\documentclass[12pt,preprint]{aastex}

\usepackage{emulateapj5}
\usepackage{graphicx}
\usepackage{fancyheadings}
\usepackage{color}

\newcommand{\gae}{\mathrel{\raise .4ex\hbox{\rlap{$>$}\lower 1.2ex\hbox{$\sim$}
}
}}
\newcommand{\gps}{B2~0738+313}

\shorttitle{X-ray emission from B2~0738+313}

\shortauthors{Siemiginowska  et al.}

\begin{document}

\title{Chandra Discovery of an X-ray jet and Extended X-ray Structure in
z=0.63 quasar, B2 0738+313}

\author{Aneta Siemiginowska$^1$,
Carlo Stanghellini$^2$, Gianfranco Brunetti$^3$, Fabrizio
Fiore$^{4}$, Thomas L. Aldcroft$^1$, Jill Bechtold$^5$, Martin
Elvis$^1$ Stephen S. Murray$^1$ L.A. Antonelli$^4$ \& S. Colafrancesco$^4$}
\affil{$^1$ Harvard-Smithsonian Center for Astrophysics \\
$^2$ IRA-CNR, Noto, Italy\\
$^3$ IRA-CNR Bologna, Italy\\ 
$^4$ INAF - Osservatorio Astronomico di Roma, Italy\\
$^5$ Steward Observatory, University of Arizona}

\email{asiemiginowska@cfa.harvard.edu}

\begin{abstract}

We have made a $\sim 30$~ksec {\it Chandra} observation of the
redshift z=0.63 GPS quasar B2 0738+313. We detected X-ray emission
from the core and have discovered a 200~kpc (projected on the sky)
X-ray jet. The X-ray jet is narrow and curves, following the extended
radio structure to the south of the quasar, and ending with a hot spot
at the southernmost part of the radio lobe. The jet has a knot at
$\sim 13$ arcsec away from the core. The knot emission is consistent
with the X-rays being created by the inverse Compton scattering of the
cosmic microwave background (CMB) photons and requires jet bulk
Lorentz factors of a few ($\Gamma_{bulk} \sim 5-7$). We discuss the
emission mechanisms that may be responsible for the jet emission.  We
present new VLA data of the core and jet, and discuss the relation
between the extended radio and X-ray emission.  Extended emission
observed in several GPS sources has been interpreted as a signature of
the source past activity, while the GPS source is young and newly
expanded.  We argue that B2~0738+313 may be an example of a new class
of radio sources similar to the FRII radio galaxies in their high jet
bulk velocities, but with the powerful GPS-like nucleus.

B2 0738+313 also has two damped Lyman-$\alpha$ systems along the line
of sight, at $z_{abs}$ = 0.0912 and 0.2212.  We discuss the possible
connection between the X-ray absorption (7.2$ \pm 0.9 \times
10^{20}$~cm$^{-2}$) detected in the ACIS spectrum and these two
intervening absorbers.  We also investigate an extended structure
within the central 10 arcsec of the core in the relation to structure
seen in the optical.

\medskip

\smallskip

\end{abstract}

\section{Introduction}

GigaHertz-Peaked Spectrum (GPS) and Compact Steep Spectrum (CSS)
sources compose a significant fraction of the bright radio source
population (10-20$\%$, O'Dea et al. 1998). However, because this
classification is based entirely on the spectral shape in the radio
band, the GPS/CSS sample is heterogeneous and contains both
galaxies and quasars with a range of luminosities and morphology.  Until
recently, most GPS/CSS sources were thought to be very compact with the
radio emission contained within $\sim1-10$~kpc of the core.  About $\sim 25
\%$ of GPS/CSS sources exhibit faint extended radio emission on
scales larger than the size of the host galaxy (Baum et al 1990,
Stanghellini et al 1990, O'Dea 1998, Fanti et al 2001) with only a few
sources showing any radio structures on Mpc scales (Schoenmakers et al
1999, Siemiginowska et al 2002, Marecki et al 2002). This extended
radio emission is amorphous or jet-like, but is usually hard to
classify because of being weak. A detection of a faint radio structure
in the vicinity of a strong GPS/CSS core requires difficult, high
dynamic range observations and careful data analysis. Even then the
extended emission may be difficult to detect.

The unprecedented sub-arcsecond resolution (Van Speybroeck et al,
1997) of the {\it Chandra} X-ray Observatory (Weisskopf et al 2002)
gives us, for the first time, the opportunity to study details of the
X-ray structures in the vicinity of AGN and quasars. Discoveries of
many X-ray jets associated with radio sources show that X-ray jets
have large bulk Lorentz factors ($\Gamma_{bulk} \sim 3-10$) at hundred
kpc distances from the nucleus.  Recent {\it Chandra} observations
(Schwartz et al 2000, Tavecchio et al. 2000; Celotti et al. 2001;
Siemiginowska et al. 2002; Brunetti et al. 2002, Sambruna et al. 2002)
have shown that if even mildly relativistic plasma is present in a jet
with high bulk velocity then it can result in a strong X-ray emission.
The cosmic microwave background (CMB) photons Compton scatter off
relatively low energy relativistic electrons ($\gamma \sim 10-300$)
within a jet creating X-rays. Both high $\Gamma_{bulk}$ and the
increased energy density of the CMB radiation (as $(1+z)^4$)
contribute to the final X-ray intensity (see Celotti et al 2001,
Harris \& Krawczynski 2002, Schwartz 2002).  Thus X-rays provide a
good indicator of the large scale environment of GPS quasars revealing
jets or diffuse cluster emission at high redshift (Siemiginowska et al
2003) allowing a differention between truly compact and ``extended''
sources within the GPS/CSS sample. The reliable separation of jet
dominated GPS/CSS sources from others in this heterogeneous class may
clarify the nature of non jet dominated GPS/CSS sources.  At the end
of Section 7.1 we discuss the similarities between B2~0738+313 and
other core dominated quasars and the implications of our observation
on the classification of this source.

B2~0738+313 (OI~363) is a redshift z=0.63 radio-loud quasar with a
peak in its radio spectrum at $\sim$5~GHz.  \gps\, is a low
polarization quasar.  The radio structure (Stanghellini et al. 1997,
2001, Kellermann et al. 1998) shows a jet on the milliarcsecond
scale. The jet is elongated towards the south for about 5 milliarcsec
($>45$ parsecs at z=0.63). It then bends by about 45 degrees in the
south-west direction, where it is visible for other 5
milliarcseconds. A bright knot is located at the jet's bend. Murphy et
al. (1993) and Stanghellini et.al. (1998) presented a VLA radio map
which shows a faint radio structure on the arcmin scale, extending
north-south in both directions from the core.

B2~0738+313 has also two intervening damped Lyman-$\alpha$ absorption
systems (DLA) along the line of sight at redshifts 0.0912 and 0.2212
(Rao \& Turnshek 1998). These large column density systems can produce
significant low energy X-ray absorption (Bechtold et al 2001, Turnshek
et al 2003). There is no Lyman-$\alpha$ absorption at the quasar
redshift.

We observed B2~0738+313 with {\it Chandra} in order to study both
extended X-ray emission indicated by the ROSAT HRI imaging, and to
search for absorption possibly related to the DLA systems.  Here we
present the results of both ROSAT and {\it Chandra} observations. In
{\it Chandra} we detect a curved X-ray jet which extends $\sim
35\arcsec$ away from the quasar. We also detect symmetric, extended
X-ray emission within 10$\arcsec$ of the quasar in the {\it Chandra}
data.

We have obtained new VLA radio data, and combined previously published
data to study B2~0738+313 at different frequencies and antenna
configurations.  We discuss the correlation between radio and X-ray
structures.  We model the radio to X-ray jet emission, and discuss the
implications for possible emission processes.  Synchrotron emission
can explain the hot spot located at the furthest distance from the
quasar core.  The emission from the knot at $\sim 13 \arcsec$ from the
nucleus requires an additional component and we argue that the
interactions between the relativistic particles and the CMB photons
are responsible for the observed X-ray flux.

We discuss briefly the X-ray spectrum of the quasar core, its
luminosity and absorption column. We consider a possibility that the
observed absorption is related to one or both of the two DLA
systems. We also compare the X-ray structure within $\sim 10 \arcsec$
from the core to optical images and discuss their possible relation to
the DLA galaxies.

We assume H$_0$=75~km~sec$^{-1}$Mpc$^{-1}$, q$_0=0.5$ (1~arcsec $\sim
5.6$~kpc).  Adopting the best fit cosmological parameters implied by
the recent WMAP results (Spergel et al. 2003)
H$_0$=71~km~sec$^{-1}$Mpc$^{-1}$, $\Omega_{\lambda}$=0.7, and
$\Omega_{M}$=0.3 changes the luminosity distance by about 10$\%$.

\section{X-ray Data}

Table~\ref{tab:date} lists the dates of the X-ray observations.  The
purpose of the ROSAT HRI ({\it R\"ontgen Satellite} High-Resolution
Imager\footnote{http://heasarc.gsfc.nasa.gov/docs/rosat/rosdocs.html})
imaging of \gps\ in 1994 was to search for the extended X-ray emission
surrounding the source. The only published results from this
observation is the conference contribution of Antonelli and Fiore
(1997). There is a hint of the extended emission in the $\sim
3\arcsec$ resolution image which prompted us to ask for the higher
resolution {\it Chandra} observations.  Here we present both the
analysis and results of the ROSAT HRI observations and the new {\it
Chandra} results.

\subsection{ROSAT HRI observations}

The ROSAT HRI observation was taken on April 5-12, 1994 with total
exposure time of 47,508 sec.  HRI source counts were accumulated in
channels 2-12, to reduce the background and possible UV contamination
and were extracted from a circle with 20 arcsec radius.  Background
was extracted in two ways: from annuli of inner and outer radii:
0.4'-0.8' and 1.7'-2.1' and from boxes of various sizes, from 20$''$
to 3$'$ on side, around the quasar. Point sources with signal to noise
ratio $>3$ were excluded from the background regions. Different
background regions gave essentially the same results. We obtain
860$\pm30$ source counts (within 20 arcsec radius circle)
corresponding to the count rate of 0.018 counts~sec$^{-1}$ with the
background level of 1.04$\times 10^{-6}$
counts~sec$^{-1}$~arcsec$^{-2}$ (e.g. 62 counts within 20 arcsec
radius circle).

The ROSAT HRI observations were obtained over the span of a few days.
There are groups of contiguous data (several OBIs, Observation
Intervals) separated by gaps of 0.5-1 days.  To search for systematic
aspect-related errors we examined the HRI images from separate groups
of contiguous OBIs.  Each image contains between 80 and 200 counts.
The faintness of the quasar count rate prevents a finer splitting of
the data (as in Morse et al. 1995) since the error on the centroid
position is of the order of a few arcsec ($\sim 2$ arcsec) for images
with about 80 counts.  We found no evidence for a systematic shift of
the image centroids between each data set and therefore no artificial
broadening in the final image.  The final location of the centroid is
offset from the optical position by 1.9 arcsec which is within a
typical accuracy ($\sim 2$ arcsec) of the source location obtained
with ROSAT HRI.

\begin{figure*}
\epsscale{0.85}
\plottwo{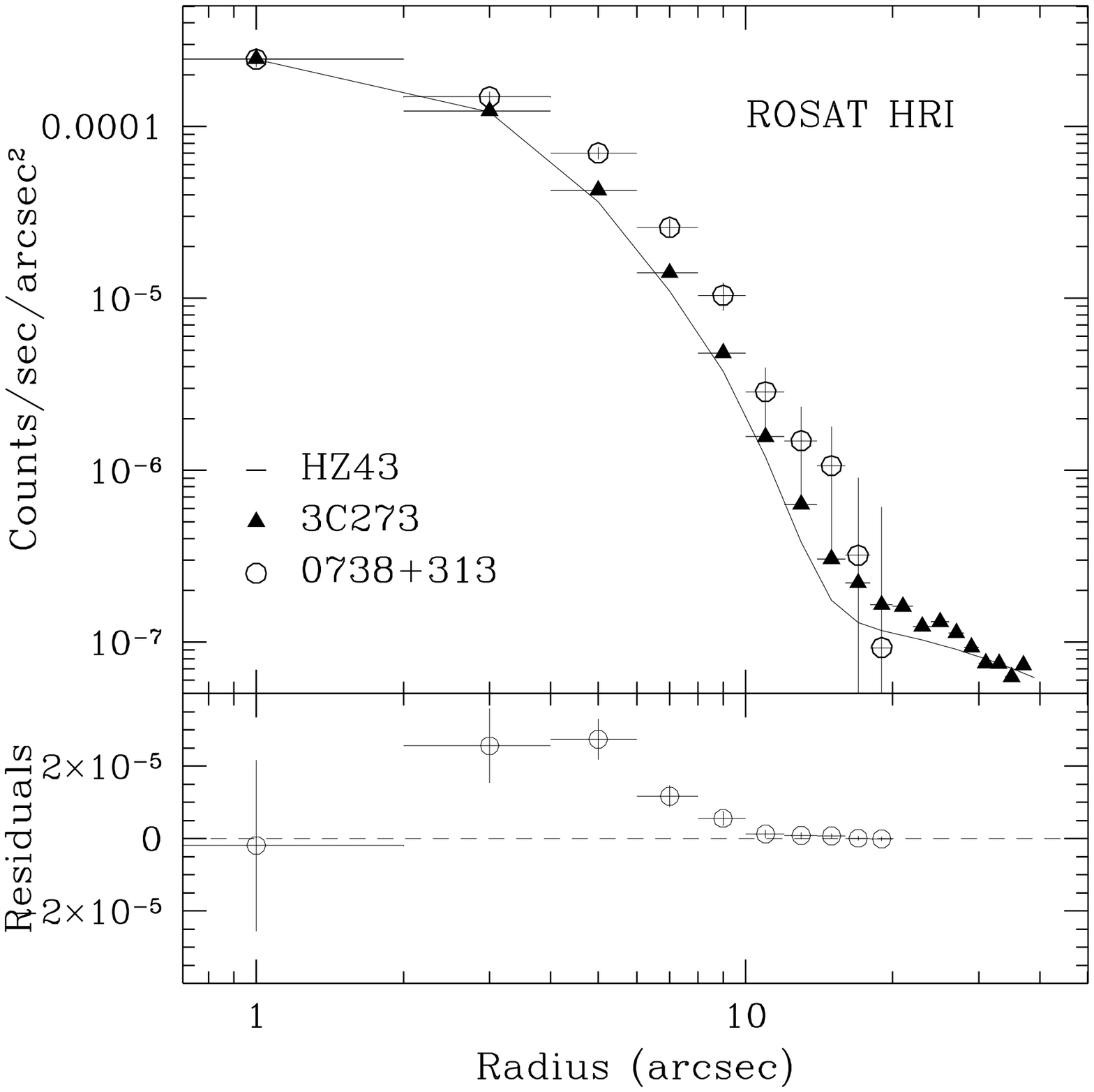}{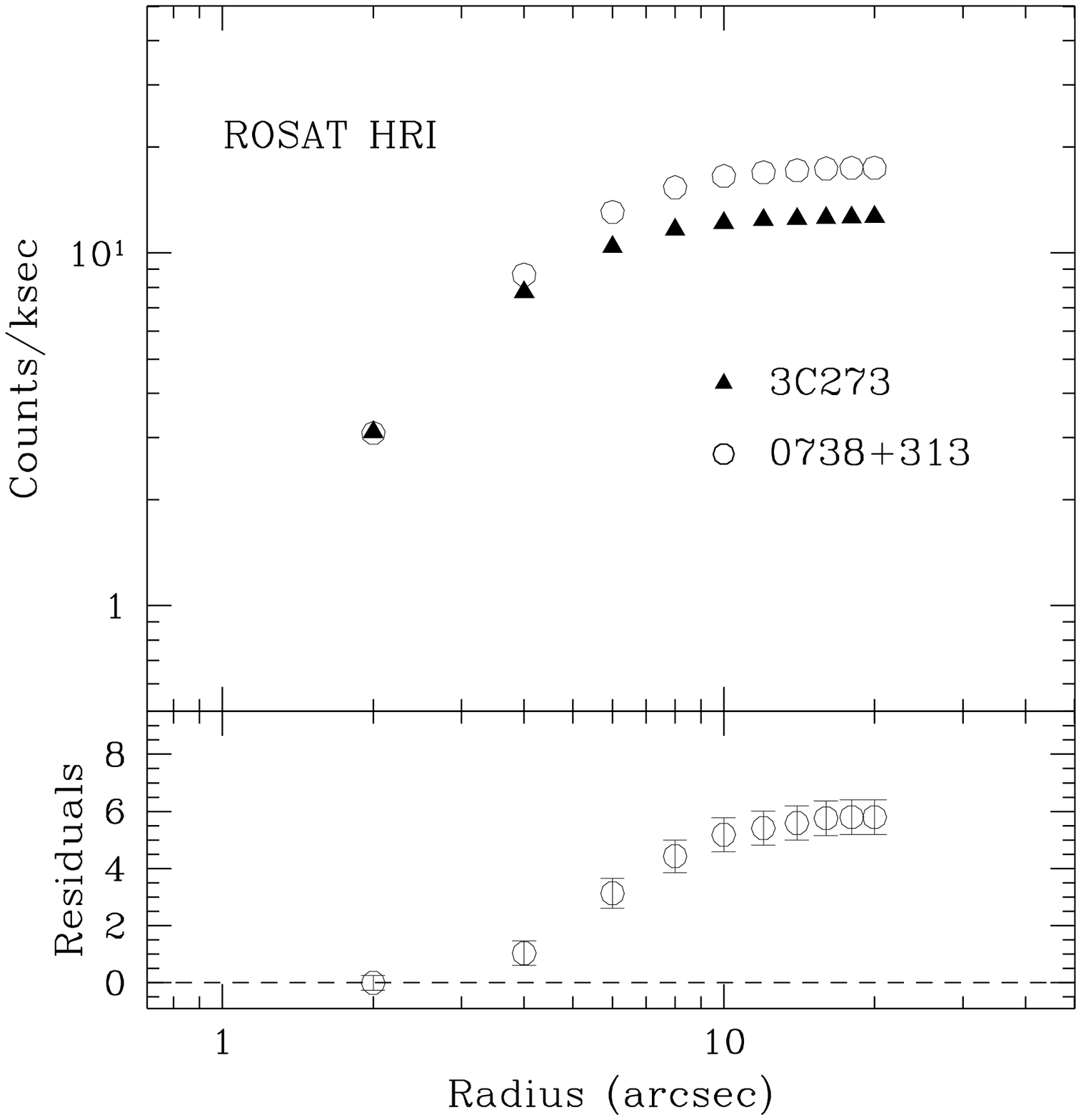}
\caption{(a) The ROSAT HRI radial profile of B20738+313
compared with that of 3C273 (after removing the jet region) and with
the David et al HZ43 HRI PSF. (b) The residual plot of B2~0738+313 and
3C273 profiles: upper panel shows the integrated counts/ksec for a
given radial distance from the center. Lower panel shows the
difference between the B2~0738+313 profile and the 3C273 profile.}
\label{fig:hri}
\end{figure*}

To check for residual spurious broadening of the HRI PSF we analyzed
the observation of the bright quasar 3C273. 3C273 is a point-like
source except for the jet emission and therefore can be used to
evaluate the parameterization of the HRI PSF.  In
figure~\ref{fig:hri}a we show the radial profile of B2~0738+313
compared with that of 3C273 (after removing the jet region) and with
the HRI PSF, as parameterized by David et al (1993).  The profile of
3C273 differs slightly from the HRI PSF beyond about 5 arcsec radius.
This broadening is likely to be due to residual systematic
uncertainties in the aspect solution as a function of time, as the
satellite wobbled back and forth by several arcmin.  The excess of
counts in 3C273 with respect to the PSF is $\sim$ 20$\%$ between 5 and
10 arcsec and $\sim$ 45$\%$ between 10 and 20 arcsec.  For B2~0738+313
the excess of counts with respect to the PSF is $110\pm15
\%$ between 5 and 10 arcsec and $190\pm70\%$ between 10 and 20 arcsec.
We therefore conclude that significant extended emission is present in
B20738+313 on scales from 5 to about 20 arcsec, where the source
starts to become fainter than the background.  This is shown more
clearly in the residual plot of Figure~\ref{fig:hri}b, where we plot
integrated counts at a given radius away from the center for both
B2~0738+313 and 3C273 and show the difference between both
profiles. We estimated the count rate in the extended emission to be
within 4.8-5.8 counts~ksec$^{-1}$, which is about 30$\%$ of the total
quasar count rate. The observed 0.1-2~keV flux in the extended
component is of order 1.6$\times$10$^{-13}$ergs~cm$^2$~sec$^{-1}$,
which corresponds to the rest frame luminosity of 1.5$\times
10^{44}$ergs~sec$^{-1}$.

The {\it Chandra} observation described in the next section allowed
further investigation of the extended emission seen in the HRI data.

\subsection{Chandra ACIS-S Observations}

We observed \gps\ for 27,600 seconds with the spectroscopic array of
the {\it Chandra} Advanced CCD Imaging Spectrometer (ACIS-S, Wiesskopf
et al 2002) on 2000 October 10 (ObsID 377) without any transmission
grating in place.  The source was located on the back illuminated chip
(S3) and offset by $\sim 35\arcsec$ from the default aim point
position to avoid the node boundary ({\it Chandra} Proposers'
Observatory Guide, POG, 2000).  {\it Chandra} ACIS-S3 data were
reprocessed with the pipeline version 6.0.1 on June 13, 2001.  We have
used the reprocessed data and calibration files available in CALDB
v.2.9. We note that the aspect uncertainty on the absolute position is
less than $\sim 0.5\arcsec$ and the uncertainty of the aspect solution
is less than $\sim 0.1\arcsec$ (Aldcroft et al 2000). The PSF FWHM at
the quasar core is about $0.75\arcsec$.  The X-ray position of the
quasar (J2000: 07:41:10.7 +31:12:00.2) agrees with the optical
position (Johnston et al 1995) to better than 1$\arcsec$, as expected
given the quality of the {\it Chandra} aspect solution (Aldcroft et al
2000).  We used CIAO (version 2.2
\footnote{http://asc.harvard.edu/ciao/}) software to analyze the data.

The extended emission detected with the ROSAT HRI observations is
resolved into smaller structures in the {\it Chandra} ACIS-S data.
Here we can clearly identify the bright quasar, a few features within
3-10 arcsec from the quasar and the curved jet extending up to $\sim
35\arcsec$ into the South-East.

\begin{figure*}
\epsscale{0.75}
\plotone{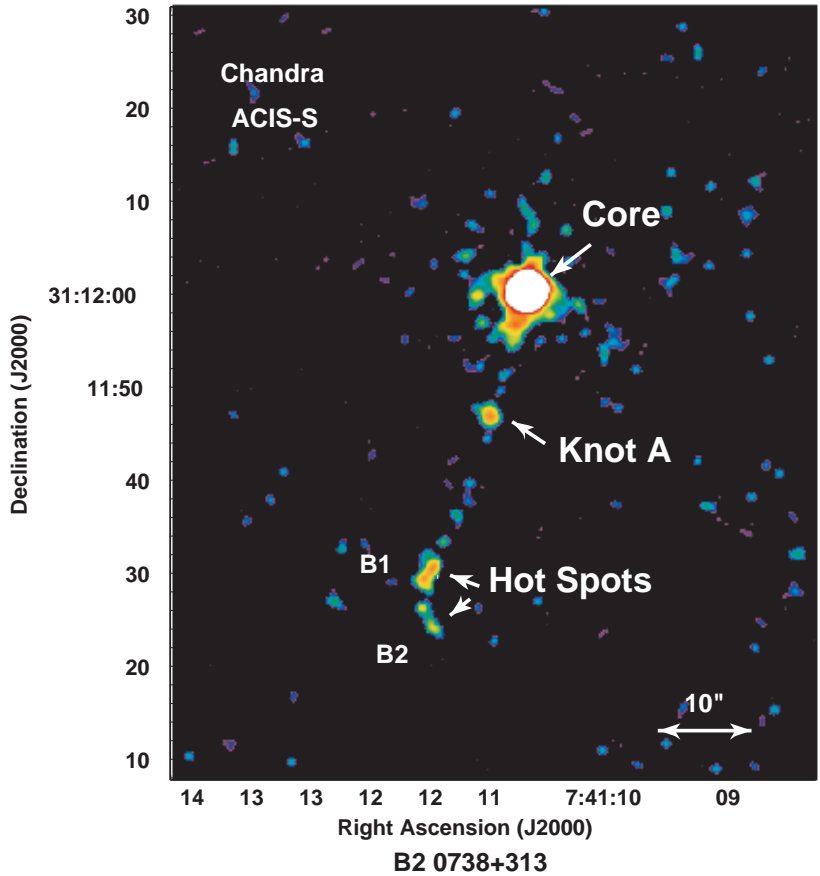}
%\plotone{acis-grid-cmyk.ps}
\caption{\small \gps\, {\it Chandra} ACIS-S image smoothed 
with a Gaussian kernel (FWHM= 0.75$\arcsec$).  Because ACIS-S
background is increasing rapidly at high energies only the events with
energies between 0.3-6.5~keV were included in this image. The jet
components, A,B1 and B2 are indicated in the figure. North is up and
East is left. The color scale corresponds to the surface brightness of
$<5 \times 10^{-8}$ in purple, $0.5-1.0 \times 10^{-7}$  dark to light
blue, $1.5 - 2.5 \times 10^{-7}$ green, $2.5 - 4 \times 10^{-7}$
yellow/orange, $4-8 \times
10^{-7}$photons~cm$^{-2}$~sec$^{-1}$arcsec$^{-2}$ in red. The white
core is $> 10^{-6}$ with a maximum at 1.3$\times 10^{-4}$
photons~cm$^{-2}$~sec$^{-1}$arcsec$^{-2}$  
The 10 arcsec scale bar is shown in right corner.}
\label{fig:acis}
\end{figure*}

A smoothed image of \gps\ in 0.3-6.5~keV is shown in
Figure~\ref{fig:acis}. The original event file was binned (into 1/3 of
ACIS-S pixel size = 0.164$\arcsec$), divided by the exposure map to
obtain a surface brightness image and then smoothed with the Gaussian
kernel (FWHM=0.75$\arcsec$).  The intensity in the image
(Figure~\ref{fig:acis}) has been scaled logarithmically to emphasize
the faint jet emission.  The small {\it Chandra} Point Spread Function
(PSF), especially the low power in the scattering wings
($r>1\arcsec$), allows for a high dynamic range image, which is
essential to the detection and resolution of the extended structure.
A total of 3,691 counts was detected in the quasar core (for a
2.7~arcsec radius of the circular extraction region) and 124 counts in
the jet (assuming a polygon region along the jet). Thus, the total jet
emission is $\sim 30$ times fainter than the core and the individual
components in the jet are $\sim 100$ times fainter than the core.

\section{VLA Data and Radio Morphology}

\gps\, has been observed at the VLA \footnote{The
National Radio Astronomy Observatory is a facility of the National
Science Foundation operated under cooperative agreement by Associated
Universities, Inc.} with various configurations and frequencies (see
Table~\ref{tab:date}).  The B configuration data at 1.36 and 1.66~GHz
were taken during 2 observing sessions on August 22, 1998 and
September 21, 1998. We did not use September 21, 1998 data in the
analysis, because they were of poor quality due to strong
interferences across the observing band.  The C configuration data
were taken on July 26, 1997 at 4.5, 4.9, 8.0 and 8.4~GHz. Two or three
$\sim$5 minutes snapshots have been obtained at different hour angles
for each data set to improve the UV coverage.  To improve the
resolution at low frequency we also made use of a short $\sim$10
minutes observation on August 10, 1999, at 1.36 GHz in A
configuration.  Calibration, self-calibration and imaging has been
done in AIPS following the standard procedures. To improve the
resolution and to better disentangle the more compact emitting regions
from the diffuse emission, while preserving the good sensitivity the
images at 1.36~GHz and 1.66~GHz were made with a tempered uniform
weighting (ROBUST 0 in task IMAGR in AIPS), while at other frequencies
(not shown) they were made with a natural weighting to maximize the
sensitivity.

High dynamic range data are necessary to measure the low brightness
structure in the presence of a very strong component.  After several
iterations of self calibration and clean/restore procedure, we have
achieved an r.m.s. noise close to the thermal noise ($< 0.1$mJy) in
most of the field-of-view. This corresponds to a peak to r.m.s. noise
ratio of $>$20000, but the removal of the side lobes is less accurate
close to the strong core emission ($\sim 2$Jy at 1.3/1.6 GHz, $\sim 4$
Jy at 4.7/8.2 GHZ),  resulting in confusion between the true
emission and some residual side lobes.  This may be particularly true
at 1.66~GHz which is known to be a frequency plagued by interference
at the VLA. In addition it is hard to separate the contributions of
the compact components from the diffuse lobe emission.

\begin{figure*}
\epsscale{0.85}
\plotone{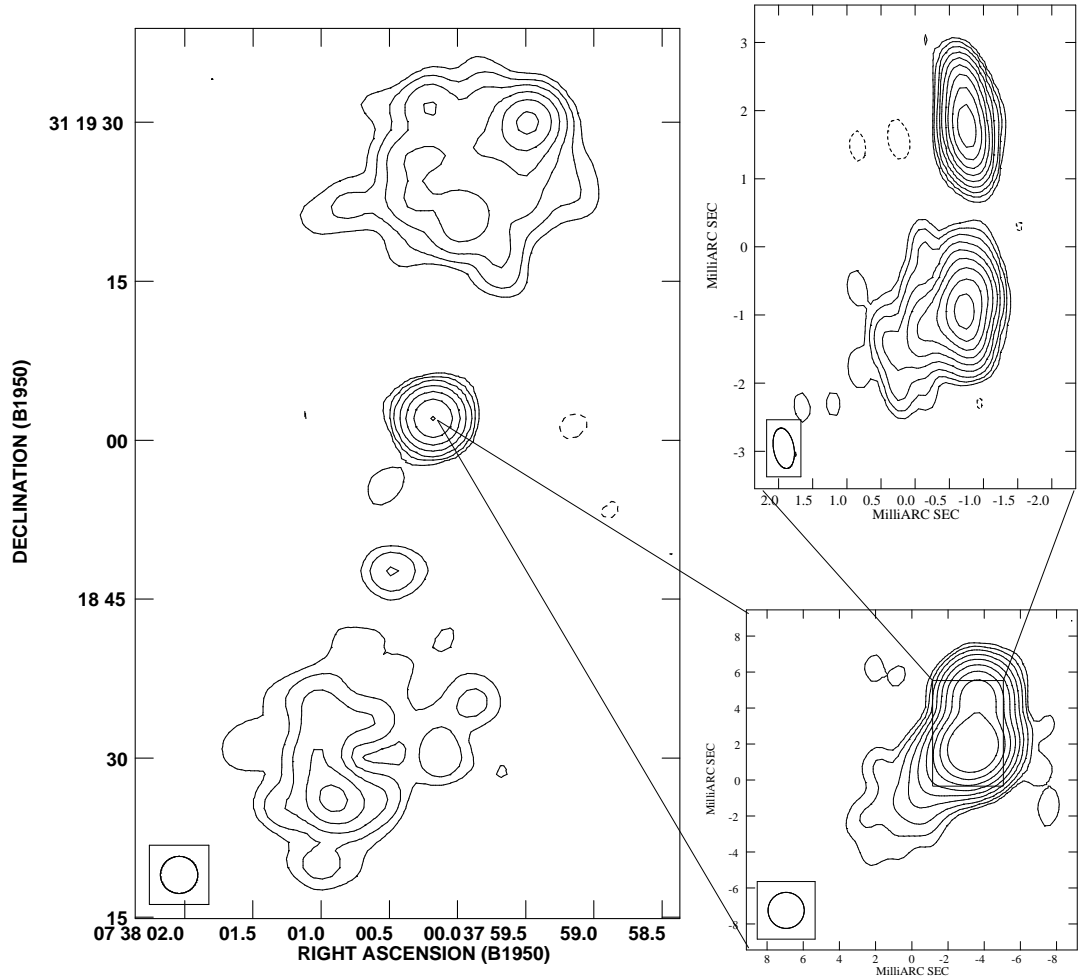}
\caption{Radio morphology of B2~0738+313 from milliarcsec to arcsecond scale.
Top right: VLBA image at 15 GHz. Bottom right: global VLBI image at 5
GHz. Left: VLA B configuration image at 1.4 GHz, with most of the core
emission subtracted. The rms noise on the images are 0.5, 0.7, and 0.1
mJy for the 15 GHz, 5 GHz, and 1.4 GHz images respectively.  The level
contours in all the images are -3, 3, 6, 12, 25, 50, 100, 200, 400,
800 times the rms noise. The beam is indicated in the bottom left
corner of each image: VLA -3.5x3.5 arcsecs, VLBI 15 GHz: 0.6x0.3mas at
PA +10 and VLBI 5~GHz: 2x2 mas.  The VLA data are discussed in the
text, the VLBI images have been produced using the data presented in
Stanghellini et al. (1997, 2001).}
\label{fig:radio-com}
\end{figure*}

\begin{figure}
\epsscale{0.75}
\plotone{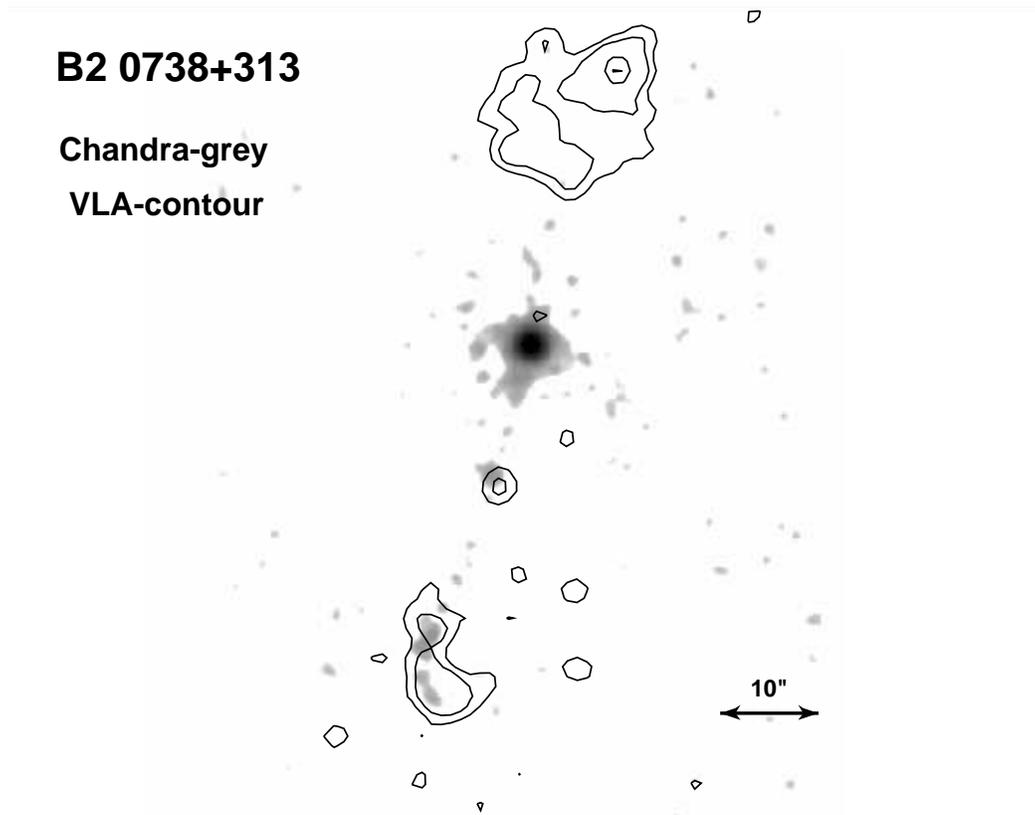}
\caption{Superposition of B2~0738+313 {\it Chandra} X-ray image
(grey) with the VLA high resolution radio (1.4 GHz) contours. The
strong radio core has been subtracted. Contour peak radio flux is at
9.4$\times 10^{-3}$ Jy/beam and and contour levels are
(3,5,11,55,75,90)$\times10^{-4}$ Jy/beam.  Grey scale in the X-ray image
varies from 10$^{-7}$ to 1.2$\times 10^{-4}$
photons~cm$^{-2}$~sec$^{-1}$~arcsec$^{-2}$.
(1 pixel=0.164 arcsec). North is up and East is left.}
\label{fig:overlay}
\end{figure}

We have made images at different resolutions cutting or tapering the
longest baselines in order to enhance different details, which
resulted in roughly circular fitted beams of the size indicated below.
In Fig.\ref{fig:radio-com}(left) we show the 1.36 GHz image of
\gps\ from the August 1998 data, with a resolution of $3.5\times 3.5$
arcseconds where the FRII type large scale morphology is well evident.
After subtraction of the variable core radio emission we combined the
B configuration data at 1.36 GHz of August 1998, with the A
configuration data at 1.36 GHz taken on August 1999 to get closer to
the Chandra angular resolution.  Then we obtained an image of
$2.5\times 2.5$ arcseconds resolution, shown superimposed onto the
X-ray image in Fig.\ref{fig:overlay}.

The image at 4.7 GHz, where the dominant component has a flat
spectrum, has been obtained from the combined UV data sets at 4.5 and
4.9 GHz, while the data at 8.2 GHz have been obtained averaging the
8.0 and 8.4~GHz data. The combined images were used to obtain the flux
density measurements of the core and extended structure.

The VLA radio images show a strong, dominant core (Table 3, partially
subtracted in Fig.\ref{fig:radio-com} to expose emission close to the
core) and a low brightness double-lobed structure extending
approximately 1 arcmin in the north-south direction. Two relatively
compact components are easily seen at the edges of the two radio lobes
and can be interpreted as the two hot spots.  The diffuse emission
around the two hot-spots is due to the lobe emission and a possible
contribution from the jet.  Previous VLA A configuration observations
(Stanghellini et al. 1998) did not reveal the extended lobes because
they were shorter snapshots and because the longer baselines of the A
configuration are less sensitive to extended emission.

The dominant radio core component is resolved in VLBI images into the
core-jet structure shown in Fig.\ref{fig:radio-com}(right)
(Stanghellini et al 1997, 2001) suggesting that the core emission is a
result of the inner jet emission relativistically boosted towards us.

\section{Jet Morphology and Spectra}

In Figure~\ref{fig:overlay}, we show a superposition of the X-ray and
radio images (1.36~GHz).  The resolution of the radio data is worse
than the resolution of {\it Chandra} at this frequency.  The X-ray jet
follows the radio structure and curves when it reaches the end of the
southern radio lobe. The enhancement between the core and the southern
components can be identified as knot A and is seen in both radio and
X-ray images.

There is no X-ray counterpart to the northern radio lobe at the level
of 7.4$\times 10^{-8}$~photons~cm$^{-2}$~sec$^{-1}$~arcsec$^{-2}$
(3$\sigma$ limit).

\begin{figure*}
\epsscale{0.85}
\plotone{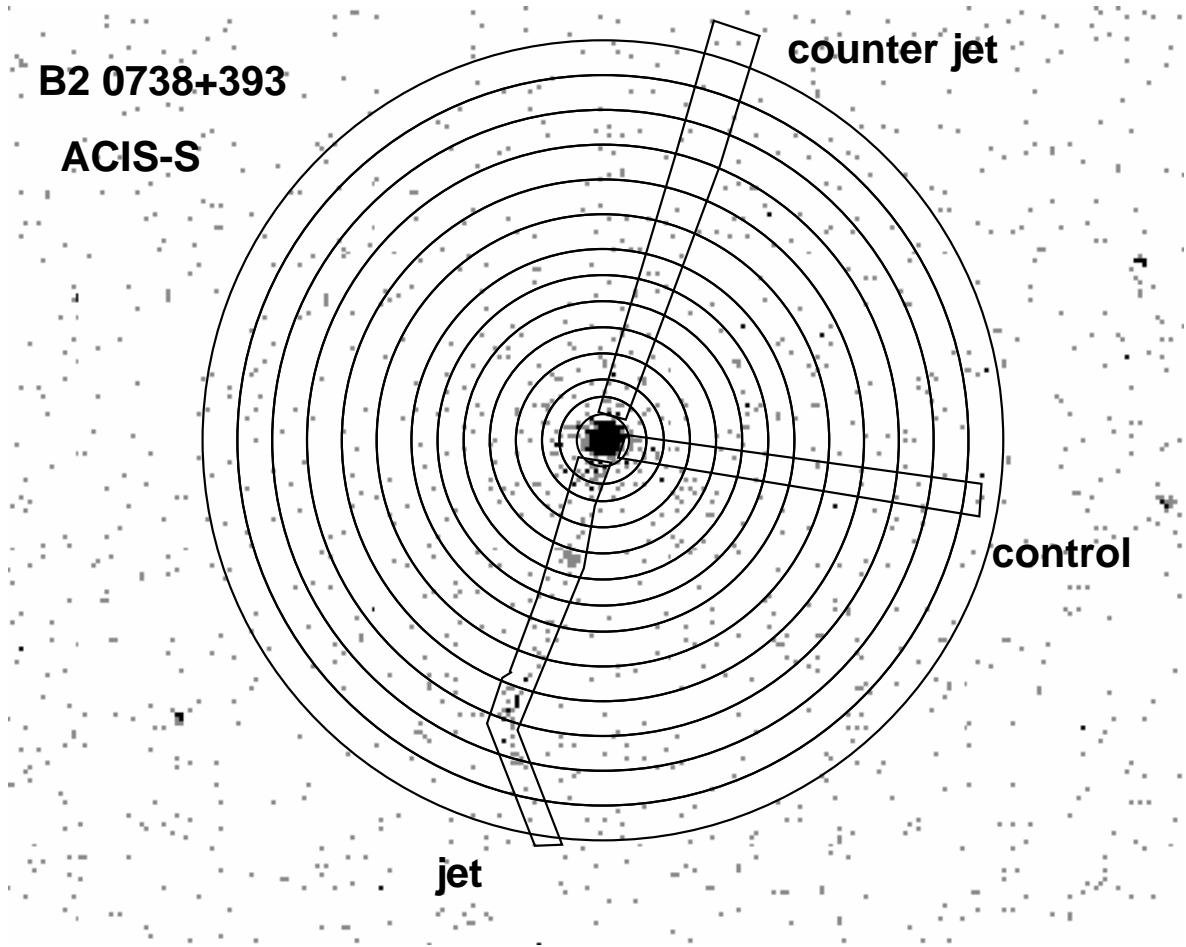}
\caption{\small Regions overlayed on the {\it Chandra} ACIS-S image
(grey scale). Only events between 0.3-10~keV are present in the image.
Intersections between annuli and polygon regions for jet, counter-jet
and control data are shown in the image.}
\label{fig:jet_regions}
\end{figure*}

\begin{figure*}
\epsscale{0.85}
\plottwo{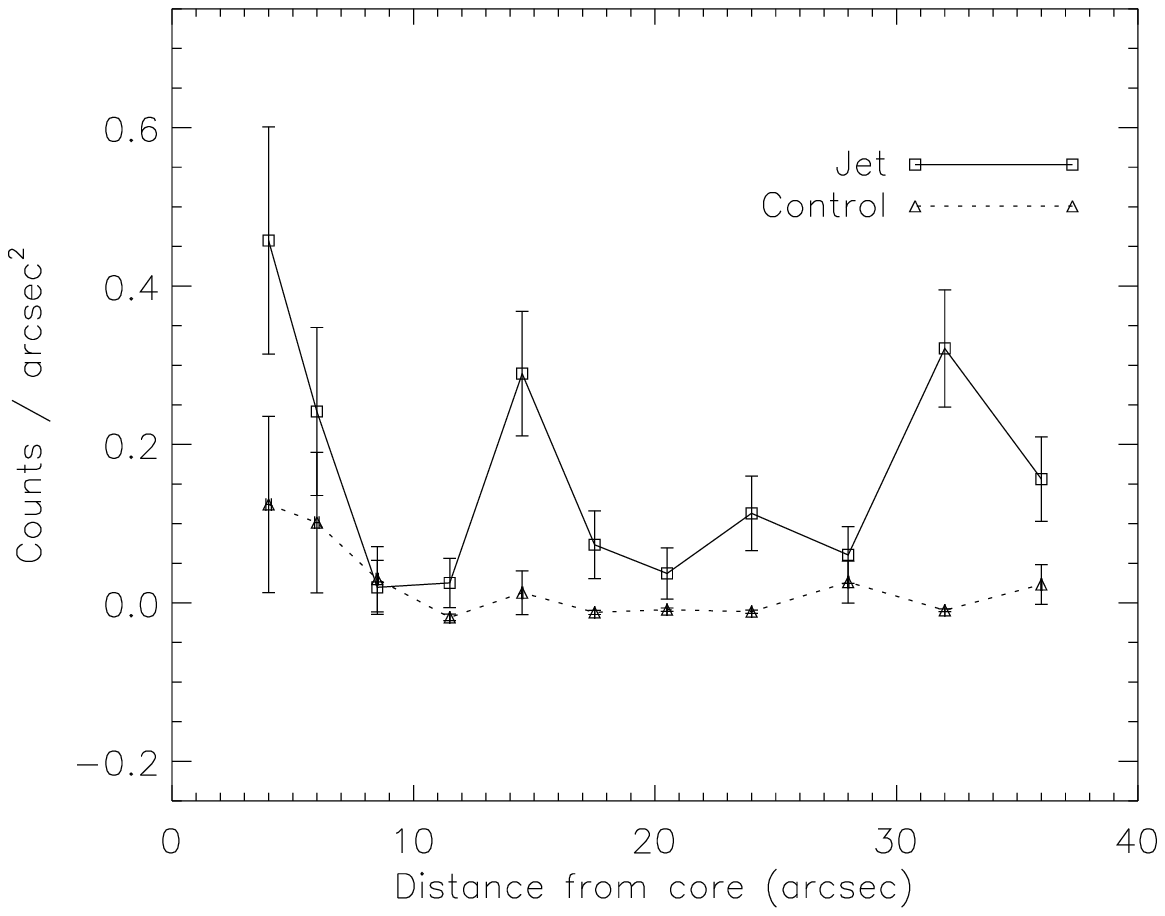}{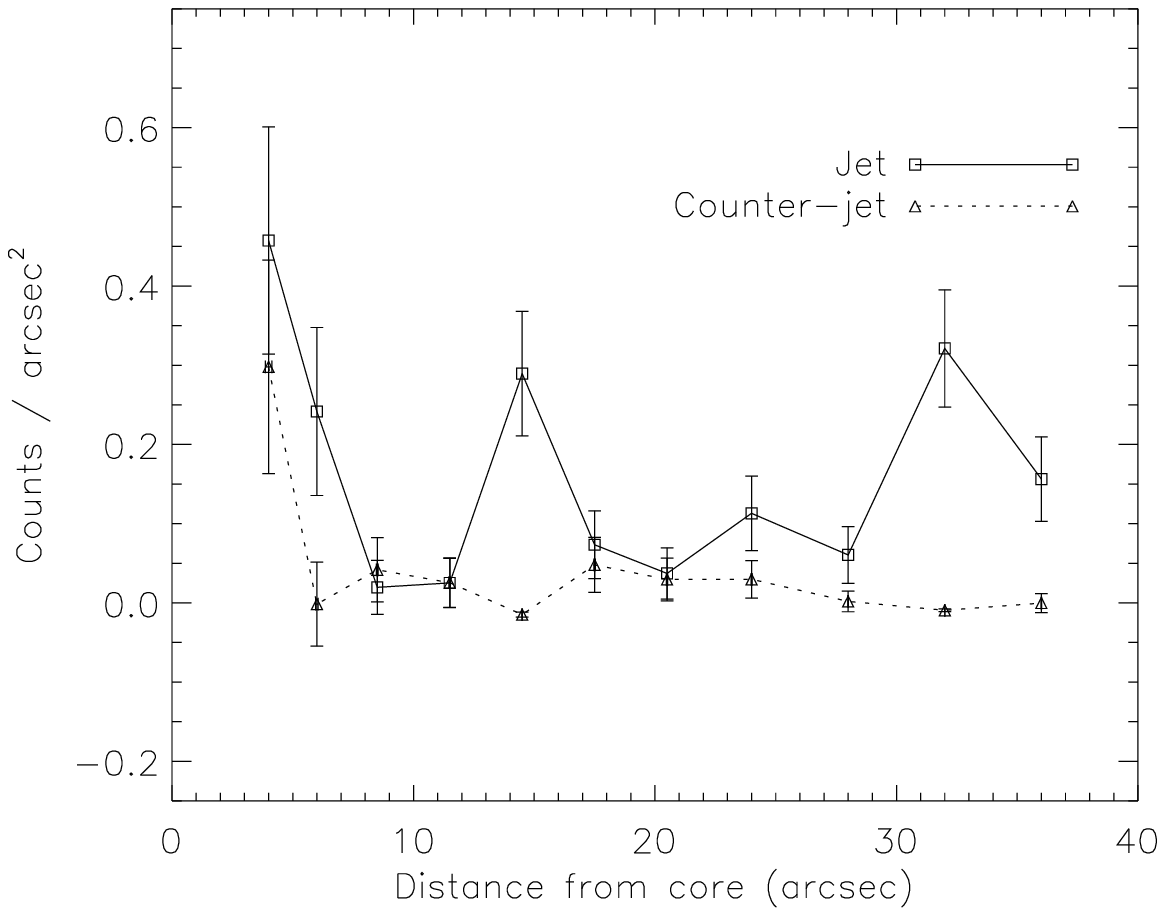}
\caption{\small Radial profiles: a) jet and the control region; 
b) jet and counter-jet.}
\label{fig:profile}
\end{figure*}

To investigate the jet structure and X-ray spectra we extracted the
X-ray jet and counter-jet profiles from the event file including only
the photons with energies between 0.3 and 6.5 keV. The {\it Chandra}
background increases significantly above 6.5 keV and below 0.3~keV
\footnote{http://cxc.harvard.edu/cal/Acis/Cal\_prods/bkgrnd/current/background.
html}.  We defined the polygonal regions shown in
Figure~\ref{fig:jet_regions} for extracting the profiles.  The
profiles are shown in Figure~\ref{fig:profile}. There is no
counter-jet present in our X-ray data.

The jet profile is not smooth and uniform, but shows definite features
$\sim$13$\arcsec$ and 32$\arcsec$ away from the core. The features are
labeled A,B1,B2 moving away from the core (see
Figure~\ref{fig:acis}). The first feature A can be related to the knot
visible in radio emission, while the outer B1 and B2 features can be
associated with the Southern Lobe and may mark the position of hot
spots.  We extracted the counts from these three features (A,B1,B2)
and fit a simple power law model to estimate the spectral index and
the luminosity associated with each feature. The results are presented
in Table~\ref{tab:xray}.  B1 component is the strongest. It seems
that both B1 and B2 components are softer ($\Gamma = 2.01 \pm0.33$ and
$\Gamma = 2.4 \pm0.5$ respectively) than knot A ($\Gamma = 1.54
\pm0.37$) located closer to the core.  The luminosity of each feature
is listed in Table~\ref{tab:xray} and they range between $1-4.1\times
10^{42}$erg~sec$^{-1}$. Figure\ref{fig:jetmodel1} show the spectral
energy distribution for each feature.

We fit all the radio components with Gaussian models using JMFIT in
AIPS (see Table 3). We attempted to match the components close to the
X-ray emission. The knot A radio emission is not well constrained with
the current data (uncertainties of order 10$\%$), but a slight shift
between X-ray and radio knot emission can be noticed. With the current
data we cannot really quantify the amount of shift.

The estimated radio flux densities are reliable for both northern and
southern hot spots, while the flux densities for the subcomponents of
the southern hot spot and the knot A are more ambiguous. These
features are definitely detected in our radio observations, but their
flux density cannot be determined accurately with the current
data.

\section{X-ray Spectrum of B2~0738+313}

We have extracted the ACIS-S spectrum of the quasar core assuming the
circular region with 2.7 arcsec radius for a source and an annulus
between 5$\arcsec$ and 55$\arcsec$ (excluding the source and the jet)
for the background. The total of 3691 counts in the spectrum were then
fit in Sherpa (Freeman et al 2001). The spectrum is not significantly
affected by pileup ($<10\%$). We included only the counts with
energies between 0.3-6.5~keV.  The background is significantly higher
outside this energy range.

We assume the absorbed power law model to fit the data:

$N(E)= A E^{-\Gamma}*\rm exp^{- N^{gal}_H \sigma (E) - N^{z_{abs}}_H
\sigma(E(1+z_{abs}))}$
~photons~cm$^{-2}$~sec$^{-1}$~keV$^{-1}$. 

In this formula $A$ is the normalization at 1~keV and $\Gamma$ is the
photon index of the power law.  We have assumed two components for the
absorption. The first component is related to the effective Galactic
absorption characterized by the equivalent neutral hydrogen column
N$^{gal}_H$= 4.18$\times 10^{20}$ atoms~cm$^{-2}$ (COLDEN
\footnote{http://cxc.harvard.edu/toolkit/colden.jsp}, which has column 
densities accurate to 5$\%$ Elvis, Wilkes \& Lockman 1989). This
absorption was constant during fitting. The second component is due to
the intervening absorber located at redshift $z_{abs}$, with
N$^{z_{abs}}_H$ as the equivalent hydrogen column. $\sigma (E)$ and
$\sigma E(1+z_{abs})$ are the corresponding absorption cross sections
(Morrison \& McCammon 1983, Wilms, Allen and McCray 2000). We used
Powell optimization with Primini statistics (Kearns et al. 1995) to
determine the best fit parameter values.

The best fit parameters are listed in Table~\ref{tab:xray-fits} and
the spectrum and residuals are presented in
Figure~\ref{fig:corespec}. The structure in the residuals is due to
the uncertainties in the {\it Chandra} calibration, in particular any
spectral features at $\sim 2$~keV are related to the calibration
uncertainties across the Si and Ir edges (CXC
Calibration\footnote{http://cxc.harvard.edu/ciao/caveats/}). The
residuals below 1~keV are mostly caused by the uncertainties in the
calibration of the time-dependent effects related to the ACIS
contamination layer.  This contamination layer affects modeling of the
absorption in our source.  To account for the contaminant we have
applied the most recent correction file ({\tt apply\_acisabs
v.1.1-2}\footnote{http://cxc.harvard.edu/ciao/threads/sherpa\_acisabs/})
and created an updated effective area file (ARF), which we used in
modeling the data.

We find that the best fit equivalent hydrogen column for B2~0738+393
is 7.19$\pm0.89 \times 10^{20}$cm$^{-2}$, assuming solar abundances
and that the absorber is at z=0, compared to the Galactic column of
4.18$\times10^{20}$~cm$^{-2}$ (COLDEN).  Figure~\ref{fig:contour}
represents confidence contours on column density vs. power law photon
index plane for the best fit model parameters.  The line representing
the equivalent Galactic hydrogen column is outside 99.7$\%$ confidence
contour drawn around the best fit equivalent hydrogen column of
7.19$\pm0.89 \times 10^{20}$cm$^{-2}$.  Thus there is a small excess
in the absorption towards the quasar, however our signal-to-noise and
current knowledge of the instrument below 1keV make the evidence for
the excess absorption uncertain.  Note that $\sim 10\%$ pileup may
affect the total detected column. Photon index and the absorption
column are correlated. The pileup results in hardening of the spectrum
and therefore in lower fitted absorption column, thus the true column
density may be higher by a factor of $\sim$2. However, archival
XMM-{\it Newton} data agree with the measured by {\it Chandra} column
density (Siemiginowska \& Bechtold 2003).

\section{X-ray emission within 10$\arcsec$ region around the quasar}

ROSAT HRI data indicated that the extended emission is present at
distances smaller than 10$\arcsec$ away from B2~0738+393.   We studied
carefully this central 10$\arcsec$ region in the {\it Chandra} ACIS-S
image to identify the nature of the extended emission. Extended
emission could be associated with the environment of the GPS quasar or
with the intervening DLA systems.

First we ran {\tt wavdetect} (detection tool in CIAO based on
wavelets) on the central part of the image (50$\arcsec \times
50\arcsec$). The quasar and the jet features were detected.  The {\it
Chandra} PSF is very narrow (FWHM $\sim$0.75$\arcsec$ at the location
of our source) but the contribution from the PSF wings from the strong
core could affect the extended structure at distances greater than
$1\arcsec$ from the quasar. Therefore, we have used three methods to
understand the amount of the quasar contribution to the emission
within the 10$\arcsec$ circle.

(1) We excluded the core component (within a 1.5$\arcsec$ circle) and fill
the excluded regions with Poisson noise.  We then smoothed the
diffuse image assuming a Gaussian kernel. (2) We fit the image data with
the PSF created from the PSF library using {\tt mkpsf} and subtracted
the PSF contribution related to the quasar core emission. We analyzed
the residuals and also created a smoothed image of the residuals
(assuming Gaussian kernel with FWHM=0.75$\arcsec$). (3) We fit the
surface brightness with 2D Gaussian and Lorentzian models as an
approximation of a quasar PSF and  analyzed the residuals for any
significant structure.

The result in each case was the structure presented in
Figure~\ref{fig:central} with three distinct 1.3-3.6$\sigma$ features
within $\sim 6$~arcsec to the south, west and east of the quasar.  We
extracted counts from each feature assuming regions specified in the
Table~\ref{tab:xray}.  The X-ray emission is the strongest
(3.6$\sigma$) at $\sim$3.8$\arcsec$ south of the quasar core in the
direction of the jet (shown with an arrow in
Fig.~\ref{fig:central}). This emission can indicate the location of the
innermost components of the jet.

The two features seen in our Figure~\ref{fig:central} -- one to the
west and the structure to the east -- may correspond to the structures
observed in the optical by Turnshek et al.(2001; see their Fig.1).  We
extracted spectra from the west and east features.  The number of
counts and the corresponding flux for each feature are listed in the
Table~\ref{tab:xray}. The table lists also the regions used to extract
the spectra ({\tt dmextract} tool in CIAO). We calculate the X-ray
flux assuming galactic N$_H$ of 4.18$\times 10^{20}$~cm$^{-2}$ and a
power law index $\Gamma$=1.25. We obtained 5.34$\pm3.2 \times
10^{-16}$ ergs~cm$^{-2}$sec$^{-1}$ for the east feature and 3.9$\pm3.2
\times 10^{-16}$~ergs~cm$^{-2}$sec$^{-1}$ for the west components.

The location of the west and east X-ray features correspond to the location
of the optical ``arm'' and a ``jet-like'' structures, identified by
Turnshek et al (2001) with the DLA system at z=0.0912. The agreement in 
position is less than  
1~arcsec.  Assuming that both X-ray components are related to the DLA
system at z=0.0912 their X-ray (0.1-2keV) luminosity
8.5$\times10^{39}$~ergs~sec$^{-1}$ and 6.5$\times
10^{39}$~ergs~sec$^{-1}$ respectively
(H$_0=75$km~sec$^{-1}$Mpc$^{-1}$, $q_0=0.5$).

These two features may also be associated with the emission at the
quasar redshift and not related to the DLA system. They may indicate a
presence of a cluster or a group at the quasar redshift. The east and
the west X-ray emission would be then associated with individual
galaxies and indicate their X-ray luminosities of order
5.1$\times10^{41}$~ergs~sec$^{-1}$ and
3.7$\times10^{41}$~ergs~sec$^{-1}$ respectively.

ROSAT HRI data indicated an extended emission within 5-20 arcsec from
the quasar of a total (0.5-2 keV) flux of order 1.3-1.6$\times
10^{-13}$ ergs~cm$^{-2}$~sec$^{-1}$. The {\it Chandra} flux including
all the resolved structures (jet and three features) accounts only for
about 50$\%$ of the ROSAT HRI flux. This discrepancy could be related
to the difference in the PSF wings, which could affect the ROSAT HRI
measurements.

\section{Discussion}

\subsection{Jet models}

Several emission processes are potentially responsible for the jet
X-ray emission: synchrotron, synchrotron self-compton (SSC), external
inverse compton (EIC) models, and thermal emission (for review see
Harris \& Krawczynski 2002).  Thermal emission from jet knots or hot
spots is quite unlikely since the observed X-ray luminosities require
unphysical conditions.  We fit the radio-X-ray data of the knot and
two components of the southern hot spot considering only non-thermal
processes.

\begin{figure}
\epsscale{0.85}
\plottwo{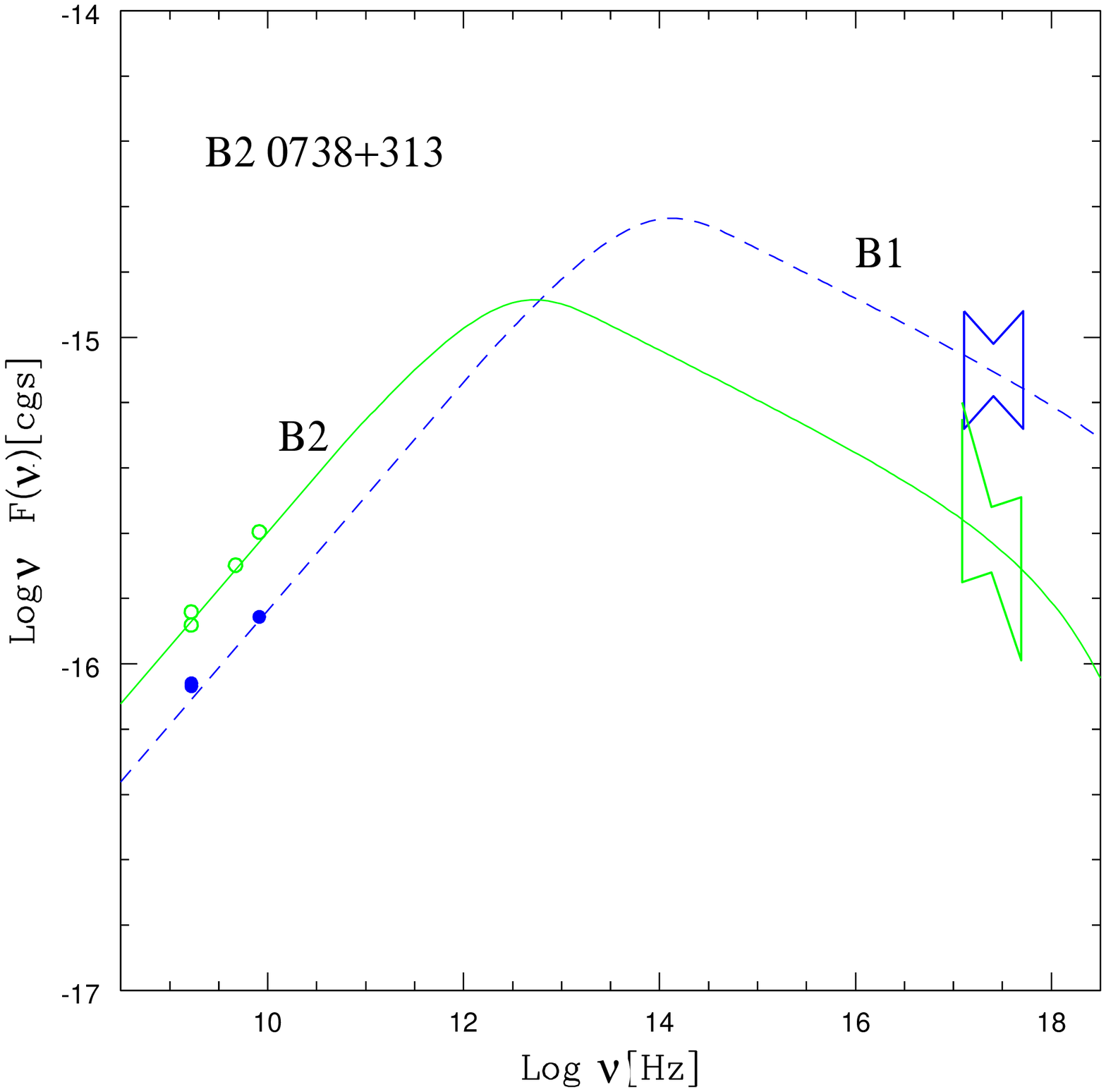}{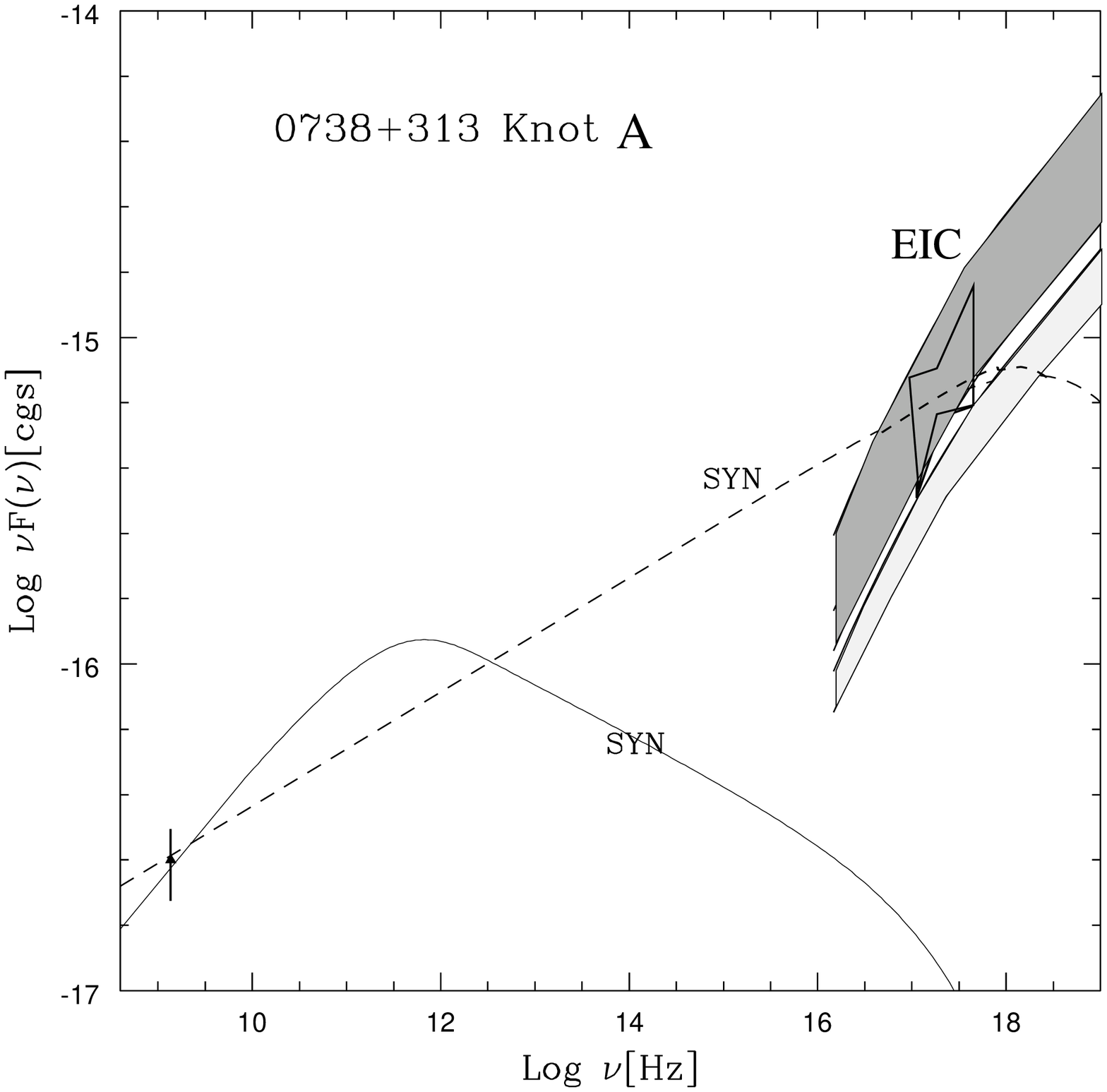}
\caption{
{\bf Left:} Synchrotron emission from the two Southernmost hot spot
components B1,B2, $\sim 30$ arcsec from the core.  {\bf Right} Spectra
predicted by different emission models for knot A: Dashed line -
synchrotron emission. Solid line indicates the synchrotron radio
component of the EIC X-ray emission.  Shaded areas show the allowed
model parameter range: dark grey: $\Gamma = 5$ for of B$<5\mu$G and
$\theta< 9 \deg$; light grey: $\Gamma = 10$ and $10<$B$<20\mu$G and
$\theta< 6 \deg$.}
\label{fig:jetmodel1}
\end{figure}

\begin{figure}
\epsscale{0.55}
\plotone{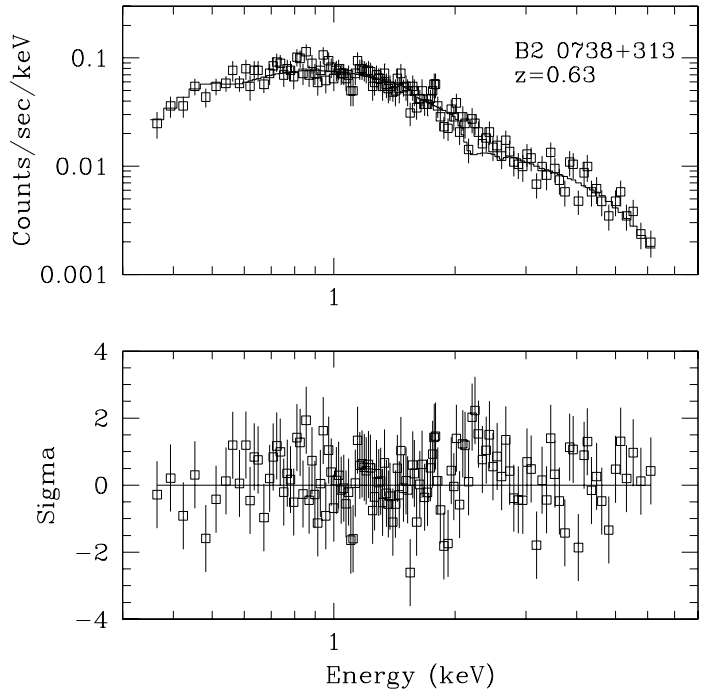}
\caption{The upper panel shows the core data fitted with the absorbed power
law model. The data were binned to 20 counts per bin for the
illustrative purpose only. Absorption is set to be at redshift 0 with
total equivalent hydrogen column of 7.2$\times 10^{20}$
atoms~cm$^{-2}$. Power law photon index $\Gamma$ = 1.55 and
normalization of
1.71$\times$10$^{-4}$photons~cm$^{-2}$~sec$^{-1}$~keV$^{-1}$ at 1~keV.
The lower panel shows the residuals in units of $\sigma$. The
variations in the residuals are due to the uncertainties in the
instrument calibration, e.g. the structure at $\sim 2$~keV is related
to the uncertainties at the Ir edge and the gain variations at the Si
edge (see {\it Chandra } POG. 2002 http://cxc.harvard.edu/proposer/POG/html/}
\label{fig:corespec}
\end{figure}

\begin{figure}
\epsscale{0.55}
\plotone{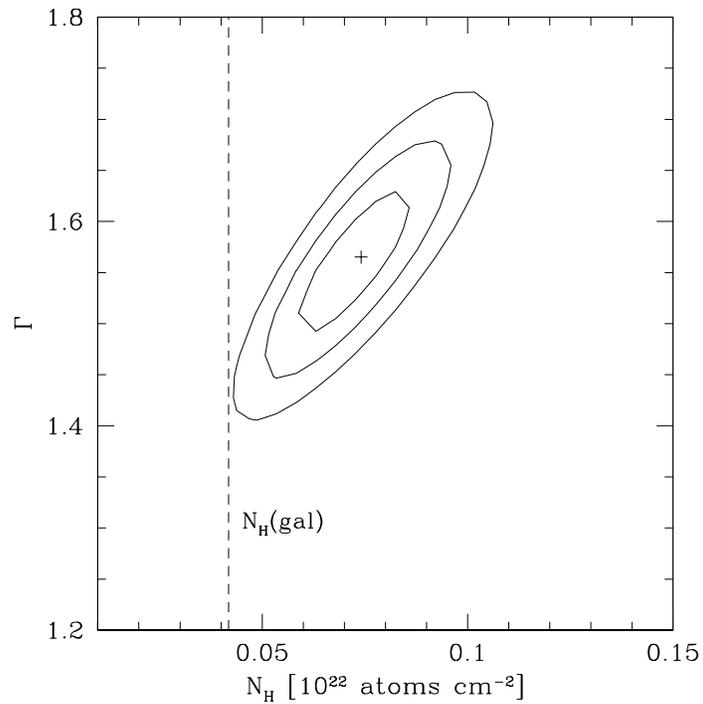}
\caption{\small Confidence regions calculated for the best fit
absorbed power law model. The best fit parameters (model A in
Table\ref{tab:xrayfits} are indicated with cross with equivalent
hydrogen column at 7.19$\times 10^{20}$~atoms~cm$^{-2}$ and $\Gamma =
1.55$. The contour represent 68.3$\%$, 95.4$\%$ and 99.73$\%$
confidence regions. The dashed line indicates the Galactic hydrogen
column of 4.18$\times 10^{20}$~atoms~cm$^{-2}$}
\label{fig:contour}
\end{figure}

\begin{figure*}
\epsscale{0.85}
\plotone{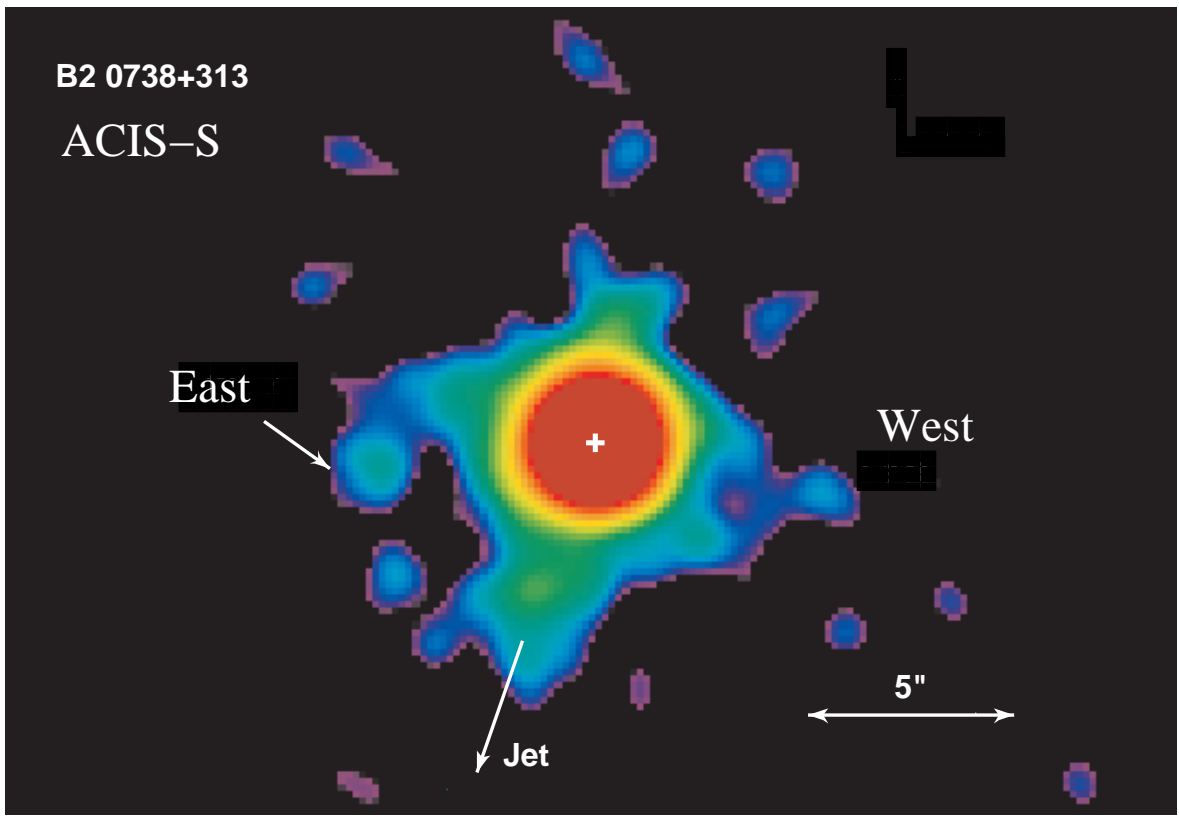}
\caption{Central 10 arcsec region.
Smoothed with the Gaussian kernel FWHM=0.75$\arcsec$. The color scale:
purple - $< 6.3 \times 10^{-8}$; blue - $0.6-1.5 \times 10^{-7}$;
green - $0.15-1.8 \times 10^{-6}$; orange: $1.8 -3.7\times 10^{-6}$
red $>5.7\times 10^{-6}$
photons~cm$^{-2}$~sec$^{-1}$~arcsec$^{-2}$. The direction of the kpc
scale jet is marked with arrow labeled Jet. The features marked West
and East may correspond to the optical features in Turnshek et al 2001
(arm and jet-like). 5 arcsec bar indicates the scale of the image and
1~pixel=0.164~arcsec }
\label{fig:central}
\end{figure*}

The X-ray emission from the two components of the hot spot in the
southern lobe (B1 and B2) is consistent with the synchrotron model
(Figure~\ref{fig:jetmodel1}). From the measured size of the radio
emitting regions and the flux density we can estimate the
equipartition magnetic field in the hot spot to be of order of
$\sim$20 $\mu$G. The radio flux symmetry between the South and North
hot spots indicate an absence of strong Doppler boosting of the radio
photons.  We can also put a rough limit on the Doppler boosting in the
hot spots using the difference in the length of the two lobes.
Assuming that the hot spot is moving at velocity $\beta = {v \over
c}$, the difference in the length (factor of about 1.3) is compatible
with an angle with the line of sight of 15-25~$\deg$ and
$\beta=0.1-0.15$.

The radio data do not provide good constraints for the synchrotron
radio emission of knot A. Figure~\ref{fig:jetmodel1} shows the radio
and data points and X-ray spectrum together with allowed models for
knot A.  The spectral break at high frequencies in synchrotron models
provides a measure of the maximum energy of the oldest plasma in the
post shock region. In general, powerful jets have spectral breaks
within $<10^{15}-10^{16}$Hz (Hardcastle et al 2002a, Sambruna et al
2002). Here this break is at $\nu_{break}>10^{19}$~Hz and implies
electron energies of at least $\gamma \sim 10^9$ (where $\gamma$ is
the Lorentz factor) with very short lifetimes ($<100$~years) for the
magnetic field expected in the jet.  However, the quality of the
present radio and X-ray data for knot A do not rule out the
synchrotron emission completely and one can find a solution with a
spectral break below $\nu_{break}<10^{19}$~Hz.

The second model for the knot A X-ray emission is Inverse Compton
scattering of cosmic-microwave background (CMB) photons off the
relativistic electrons in the jet (Figure~\ref{fig:jetmodel1}).  The
radio to X-ray luminosity ratio as well as the difference in the
spectral slopes between radio and X-ray bands requires the jet to move
with the high relativistic bulk velocities ($\Gamma_{bulk}= 5 - 10$,
where $\Gamma_{bulk}$ is the jet Lorentz factor) towards the observer
($\theta < 12$~degrees, where $\theta$ is an angle between the
observer and the direction of the jet motion). In
Figure~\ref{fig:jetmodel1} we show X-ray spectra for a range of model
parameters, $\Gamma_{bulk}$,$\theta$, $B$ - magnetic field and the
assumed slope of the energy distribution of relativistic electrons,
$\delta$. With only one radio data point for knot A the models are
unconstrained, although assuming $\delta> 2.4-2.5$ the inclination
angle $\theta$ can be up to 10-15~$\deg$. This is still in
contradiction with the morphology of the extended radio lobe
structure.

The small angle indicates that either (1) the radio lobes are $\sim
4$~Mpc in size and the jet is aligned with the lobes and very long; or
(2) that the inclination angle of the knot is different from that of
the lobes and the projected extension is of the order of
500~kpc. Because the radio lobe morphology and radio intensity suggest
a large inclination angle, while the jet requires the small
inclination angle, the second possibility may indicate either a sharp
bend in the jet or the presence of precession of the jet direction
within the last 10$^7-10^8$ years.

Megaparsec scale radio emission in GPS sources has also been
interpreted as a remnant of much earlier source activity (Baum et al
1990, Owsianik, Conway \& Polatidis 1998, Marr et al 2001;
Stanghellini et al 1990, Schoenmakers et al 1999, Marecki et al 2002).
In this model the GPS ``core'' indicates current source activity
rejuvenated by an inflow of new matter from a recent merger
(e.g. Mihos \& Hernquist 1996) or accretion disk instability
(Siemiginowska, Czerny \& Kostyunin 1996, Mineshige \& Shields
1990). Repetitive activity of the source may also come from feedback
mechanisms (Ciotti \& Ostriker 2001) in which the accreting source
heats the ambient gas to the point at which the accretion stops, since
the Compton temperature of the emitted radiation exceeds the virial
temperature of the galactic gas. Accretion restarts after the gas has
cooled. The outburst phase is short in comparison to the cooling
phase.

The above interpretation of the extended radio emission is probably
more appropriate for the sources with their sub-kpc core observed
edge-on. In such a source one can separate two jet components and the
core on the milliarcsec scale VLBI observations. For example, Compact
Symmetric Objects (CSO) which have the peaked radio spectra, but show
two jet components on both sides of the strong core have been
monitored for several years now. The dynamical timescales in these
sources can be measured and suggests young ages ($<10^4$years). The
extended, Mpc emission observed in two cases (0108+388 , Baum et al
1990 and PKS 1245+676, Marecki et al. 2002) is at least 10$^6$years
old. Owsianik et al (1998) argued based on the radio morphology that
the outer structure in 0108+388 has been left unsupplied for at least
2$\times 10^5$ years, while the core activity is only 300-400 years
old.

B2~0738+313 is not a CSO type source and we do not have any estimate
of the characteristic timescale for the activity of the core. In fact
the milliarcsec core-jet morphology may indicate that B2~0738+313
could be similar to other core dominated quasars, but with a peaked
radio spectrum. The peak in the radio spectrum could be therefore due
to the dominant contribution from the miliarcsec jet component at this
frequency. This quasar could be an example of the FRII type quasar
with a high bulk jet velocity ($\Gamma_{bulk} >5$) which has been
observed in the other FRII sources (see for example Hardcastle et al
(2002)), but with a more complex radio nuclear spectrum than usually
associated with the FRII class.

\subsection{Absorption towards \gps}

We found X-ray absorption in the core of B2~0738+313.  The X-ray
absorption may be arising in the material associated with B2~0738+313
itself, and/or one or both of two intervening damped Lyman-$\alpha$
(DLA) absorbers known, at redshifts $z=0.0912$ and $z=0.2212$ (Rao \&
Turnshek 1998).  The intervening absorbers have neutral hydrogen
columns derived from the DLA profiles of 1.5$\times 10^{21}$~cm$^{-2}$
and 8$\times 10^{20}$~cm$^{-2}$ respectively (Rao
\& Turnshek 1998). There is no Ly$\alpha$ absorption at the quasar
redshift, so it is possible that the X-ray absorption is associated
with one of the damped systems.  Unfortunately ACIS-S spectral
resolution does not allow us to obtain the redshift of the X-ray
absorption or to separate these two systems to obtain their individual
contribution to the total absorbing column.  We can only obtain the
total column of the two absorbers or fit individual columns assuming
that the absorption comes from either system (see
Table~\ref{tab:xray-fits}).  It is interesting to note that the total
equivalent column of hydrogen observed with ACIS is smaller than the
neutral column in either DLA. This suggests that the metallicity of
the absorbers may be sub-solar.  We defer a more detailed analysis of
the X-ray absorption until details of the calibration for energies
below 1 keV are well understood.

\subsection{Extended Emission}

We can use the X-ray to H-band luminosity ratios collected for a large
sample of galaxies by Shapley, Fabbiano and Eskridge (2001, SFE01) to
identify the type of galaxy possibly associated with the two X-ray
features.  Turnshek at al (2001) gives the K band magnitudes for both
components (their Table 2).  which we use to calculate the H-band
luminosity for the sources located at z=0.0912: L$_H^{arm}$=6.5$\times
10^{41}$erg~sec$^{-1}$ and L$_H^{jet-like}$=4.6$\times
10^{41}$erg~sec$^{-1}$. For such low H-band luminosity these two
features are very X-ray bright. Based on SFE01 (their Fig.6) these
features cannot be associated with the early or intermediate type of
galaxies. They might be related to late type galaxies which are
disk/arm dominated or irregular, although our X-ray luminosities are
still higher than the luminosities in SFE01 sample.  Deeper
observations in the future in both optical and X-ray domains are
needed in order to better understand the nature of these two features.

It is also possible that the west and east X-ray features are related
to the cluster or a group at the quasar redshift. If this is the case
their would have H-band to X-ray luminosity ratios typical to the
early type galaxies in SFE01 diagrams.

We do not detect an X-ray emission from the dwarf galaxy identified by
Turnshek et al (2001) with DLA system at redshift z$_{abs}$=0.2212. In
fact the X-ray emission is weakening towards the location of the
optical peak emission of the galaxy which is 5.7$\arcsec$ away from
the quasar peak.

\section{Summary}

In summary we report new X-ray and radio observations of GPS quasar
B2~0738+313.  We find that B2~0738+313
\begin{itemize}
\item is a powerful X-ray and radio source with total X-ray
luminosity $L_{0.1-10keV} \sim 10^{45}$~ergs~sec$^{-1}$;
\item has extended radio emission on 1 arcmin scale 
with FRII type morphology;
\item has a 35$\arcsec$ long, one-sided X-ray jet with high bulk
velocity,reaching the southern radio lobe.
\item has diffuse, extended X-ray emission within the central 10$\arcsec$;
\item has excess X-ray absorption on the line of sight which may be
intrinsic to the AGN, or related to one or two known intervening DLA systems.
\end{itemize}

This is the second GPS source imaged with {\it Chandra}. Like
PKS~1127-145 (Siemiginowska et al 2002) it shows complex structures
related to the radio emission on arcsec scale and absorption in the
core spectrum. The studies of GPS sources appears to be a fertile
field for understanding the evolution of AGN.

\acknowledgements

AS thanks Dan Harris for comments and discussion.  This research is
funded in part by NASA contracts NAS8-39073 and NAS8-01130. Partial
support for this work was provided by the National Aeronautics and
Space Administration through Chandra Award Number GO2-3148A issued by
the Chandra X-Ray Observatory Center, which is operated by the
Smithsonian Astrophysical Observatory for and on behalf of NASA under
contract NAS8-39073. GB acknowledge partial support by the MURST under
grant Cofin-01-02-8773.

\begin{table*}[h]
\begin{center}
\caption{Observations}
\smallskip
\begin{tabular}{lccccccc}
\hline \hline \\
Instrument     & Date     & E(keV)/$\nu$(GHz) & Exposure (s)  \\ 
\hline
ROSAT HRI & (05-12)-Apr-1994 & 0.1-4 &47508 \\
{\it Chandra} ACIS-S & 10-Oct-2000 & 0.1-10 & 27600\\
\hline
VLA (C)	 & 26-July-1997 & 4.5,4.9,8.0,8.4 & \\
VLA (B)  & 22-Aug-1998  & 1.36,1.66 & \\
VLA (B)  & 21-Sep-1998  & 1.36,1.66 &\\
VLA (A)  & 10-Aug-1999  & 1.36 \\
\hline
\end{tabular}
\label{tab:date}
\end{center}
\end{table*}

\begin{table*}[h]
\begin{scriptsize}
\caption{Measurement of the X-ray emission.}
\smallskip
\begin{tabular}{lccccccccc}
\hline \hline \\
 Component & Region & Counts & Net & $\Gamma$ & F$^a_{X}$(0.1-2)
 &L$^b_X$(0.1-2)& F$^a_{X}$(2-10) & L$^b_X$(2-10) \\
\\
\hline
Core& circle(4126.1,4023.8,5.5) & 3691 & 3674.5$^{\pm 61.2}$ &
1.55$\pm0.05$  & 346.1 & 44.52  & 868.1& 44.92\\
South & ellipse(4123.97,4015.78,3.23,3.23,0) & 21 & 17.9${\pm
4.9}$&1.27$^{\pm 0.38}$&1.42& 42.13 &5.79 & 42.75\\
East& circle(4115.23,4022.96,3.40) &12 & 6.9$^{\pm 3.6}$& 1.25$^{\pm
0.82}$& 0.53 &  41.71  & 1.88 & 42.26 \\
West& ellipse(4138.24,4022.08,6.57,2.78,0.127) &15 & 5.0$^{\pm
3.9}$&1.2$^{\pm 1.1}$ & 0.39 & 41.57 & 1.35& 42.11 \\
A & 3.6$\arcsec$ x 7.8$\arcsec$ (4118,3996)$^c$ & 30 & 21.85$^{\pm 4.8}$ & 1.54
$^{\pm 0.37}$ &
1.58 & 42.18 & 4.31 & 42.61 \\ 
B1 & 3.6$\arcsec$ x 7.8$\arcsec$ (4105,3963)$^c$ & 36 & 27.95$^{\pm 5.6}$ & 2.0
1$^{\pm
0.33}$ & 2.39  & 42.36  & 3.21 & 42.49  \\ 
B2 & 5.8$\arcsec$ x 3.5$\arcsec$ (4106,3951)$^c$ & 19 & 12.95$^{\pm 3.8}$  &
2.4$^{\pm0.5}$ &1.05 & 42.00 & 0.81 & 42.89 \\
\hline
\end{tabular}
Galactic hydrogen column of N$_H=4.18\times 10^{20}$ atoms~cm$^{-2}$
was assumed in all models; $^a$ flux within given energy band
(0.1-2~keV or 2-10~keV) in 10$^{-15}$~ergs~cm$^{-2}$sec$^{-1}$; $^b$
rest frame log luminosity within given energy range in ergs~sec$^{-1}$
assuming $q_0$=0.5, H$_0$=75~km~sec$^{-1}$Mpc$^{-1}$; $^c$ The
location of the center of the region box in {\it Chandra} physical
coordinates.
\label{tab:xray}
\end{scriptsize}
\end{table*}

\begin{table*}[h]
\begin{center}
\begin{scriptsize}
\caption{Radio Flux Density}
\begin{tabular}{lccccccccccccc}
\hline \hline \\
 Component & 1.4GHz$^a$ &     & 1.7GHz$^a$ &     & 4.7GHz$^b$&     &
8.2GHz$^b$ & & 1.4GHz$^c$ \\
	   & Peak   & Tot & Peak   & Tot & Peak  & Tot &  Peak  & Tot
& Tot \\
\hline
Core  	   & 2040   &     & 2478   &     & 4052  &     & 3942 & &\\
North HS   &13.2 & 21.5    & 9.5   & 15.9& 4.4 & 8.9 & 2.8 & 4.9 &
15.2 \\
A	   & 1.5 & 2.3     &  0.7& 1.5   & 1.1 & 4.17  & 0.6 & 1.3 & 2.\\
B1         & 3.9 & 11.27   &  2.7& 5.3   & 2.0 & 8.6   & 0.8 & 1.7 & \\
B2         & 7.7 & 12.9    &  5.5& 8.8   & 2.7 & 4.3   & 1.5 & 3.1  & 7.9\\ 
\hline
\end{tabular}
\end{scriptsize}
\end{center}

{\begin{scriptsize} 
$^a$ August 1998;
$^b$ July 1997;
$^c$ combined August 1998 and August 1999;

Flux density measurements given in mJy. The measurements for July 1997
and August 1998 JMFIT assuming a box region of a few (5 or 6) pixels
centered on the component.  The noise in the data $< 0.1$ mJy.  Total
flux density may vary by 1 or 2 mJy depending on the location of the
boxes. North HS is a radio hot spot to the North. The combined 1998
and 1999 data have much smaller measurement errors around 1~mJy at the 
hot spots and 0.2 mJy for the knot A.

\end{scriptsize}}\label{tab:radio}
\end{table*}

\begin{table*}[h]
\begin{center}
\begin{scriptsize}
\caption{Modeling the Intervening Absorption.}
\begin{tabular}{lccccc}
\hline \hline \\
 Model & $z_{abs}$$^a$ &  N$_H$($z_{abs}$)$^b$ & $\Gamma$ & Norm$^c$ & $\chi ^2
$$^d$ \\
\hline
A      &  0        & 7.19$\pm 0.90$ & 1.56$\pm0.05$ & 1.71$\pm0.07$ &
422.7(416) \\
B      &  0.09     & 3.28$\pm 1.04$ & 1.55$\pm0.05$ & 1.69$\pm0.07$ &
422.9(416) \\
C      &  0.22     & 3.85$\pm 1.19$ & 1.55$\pm 0.05$ & 1.68$\pm0.06$ & 422.6(416) \\
D      & 0.63      & 6.61$\pm 1.86$ & 1.54$\pm 0.05$ & 1.68$\pm0.06$ & 421.73(416)\\
\hline
\end{tabular}
\label{tab:xray-fits}
\end{scriptsize}
\end{center}

{\begin{scriptsize} $^a$ redshift of the absorber. Model A assumes
that the entire absorption is related to the Galaxy, while models B,C
and D assume that the Galactic column is constant at
4.18$\times$10$^{20}$~atoms~cm$^{-2}$ and the additional absorption
comes from the intervening absorber at a given, fixed redshift.  $^b$
equivalent hydrogen column of the absorber in
10$^{20}$~atoms~cm$^{-2}$ $^c$ Normalization in
10$^{-4}$photons~cm$^{-2}$~sec$^{-1}$~keV$^{-1}$ at 1~keV.  $^d$
$\chi^2$ calculated with Sherpa using Chi Primini statistics. Number
degrees of freedom is given in the brackets.
\end{scriptsize}}
\end{table*}

%%%%%%%%%%%%%%%%%%%%%%% FIGURES %%%%%%%%%%%%%%%%%%%

%%%%%%%%%%%%%%%%%%% Figure 1 ROSAT

%%%%%%%%%%%%%%%%%%% Figure 2 ACIS smoothed image

%%%%%%%%%%%%%%%%%%% Figure 3 VLA images

%%%%%%%%%%%%%%%%%%% Figure 4 Superposition VLA-Chandra

%%%%%%%%%%%%%%%%%%% Figure 5 Jet regions

%%%%%%%%%%%%%%%%%%% Figure 6 Jet profiles

%%%%%%%%%%%%%%%%%%% Figure 7 Models for the knots

%%%%%%%%%%%%%%%%%%% Figure 8 Core X-ray fit

%%%%%%%%%%%%%%%%%%% Figure 9 Contour plot

%%%%%%%%%%%%%%%%%%% Figure 10 Central image with the small structures

\end{document}